\documentclass[
    twocolumn,
	prd,
    amssymb,
	preprintnumbers,
	secnumarabic,
	nofootinbib,
	superscriptaddress]{revtex4-1}

\pdfoutput=1

\usepackage{graphicx}
\usepackage{enumitem}
\usepackage{latexsym}
\usepackage{amsfonts}
\usepackage{amssymb}
\usepackage{array}
\usepackage{pifont}
\usepackage{color}
\usepackage{xcolor}
\usepackage{amsmath}
\usepackage{slashed}
\usepackage{dcolumn}
\usepackage{verbatim}
\usepackage{float}
\usepackage{multirow}
\usepackage{xspace}
\usepackage[normalem]{ulem}
\usepackage{hyperref}
\usepackage{subfigure}
\usepackage{anyfontsize}
\usepackage{t1enc}

\definecolor{aquamarine}{rgb}{0.2,0.7,0.6}
\definecolor{cerulean}{RGB}{0,166,214} 
\definecolor{hypershade}{rgb}{0.3,0.3,0.8}
\definecolor{subtlered}{rgb}{0.8,0.3,0.3}

\hypersetup{
  pdfauthor={S Tallur},
  pdftitle={preSNnudark},
  pdfsubject={Dark matter on pre-supernova nus},
  colorlinks=true,
  citecolor=aquamarine,
  urlcolor=cerulean,
  linkcolor=gray
}

\allowdisplaybreaks

\begin{document}

\title{Dark matter scattering with pre-supernova neutrinos}

\author{Sangeetha N. Tallur}
\email{sangeethan@iisc.ac.in}

\affiliation{Centre for High Energy Physics, Indian Institute of Science, C. V. Raman Avenue, Bengaluru 560012, India}

\author{Nirmal Raj}
\email{nraj@iisc.ac.in}

\affiliation{Centre for High Energy Physics, Indian Institute of Science, C. V. Raman Avenue, Bengaluru 560012, India}


\begin{abstract}

Pre-supernova neutrinos, emitted during the last day prior to core collapse of a massive star, could provide a unique probe at MeV energies of scattering interactions between dark matter and neutrinos.
Due to attenuation of the flux of electron anti-neutrinos from their scattering on dark matter on the way to Earth,
we expect a deficit of inverse beta decay events at large-volume, low-threshold detectors such as KamLAND, JUNO, and Super-K.
which we use to derive upper limits on the dark matter-neutrino reduced scattering cross section.
Though seemingly weaker than limits from post-bounce supernova neutrino events, these sensitivities provide an important cross-check, could help determine the energy dependence of the cross section, and may even be a distinct probe of certain models.
Further, pre-supernova neutrinos may test hints reported of dark matter-neutrino scattering in the early universe suppressing small-scale power as seen in Lyman-alpha data.

\end{abstract}

\maketitle

\section{Introduction}

Observing the scattering of dark matter (DM) on Standard Model (SM) states may reveal its long-sought particle identity.
Extensive constraints have been placed on nuclear and electron recoils induced by dark matter in direct detection experiments and other terrestrial settings~\cite{Billard:2021uyg,Bramante:2026wzh},
celestial bodies~\cite{Bramante:2023djs,Mack:2007xj}, and cosmic rays~\cite{Cyburt:2002uw,Cappiello:2018hsu}; see also the references in Ref.~\cite{Strumia2003}.
In this work we consider the possibility that the dominant SM portal to DM is neutrinos, as realized in various models~\cite{Olivares-DelCampo:2017feq,Blennow:2019fhy}, and estimate the sensitivities to DM--$\nu$ scattering from a future measurement of a {\em pre}-supernova flux of neutrinos.
 
Neutrinos from core-collapse supernovae (CCSNe) have been used to set limits on interactions with DM and cosmic relics.
During their diffusion out of the proto-neutron star they can scatter with dark states produced in the supernova, prolonging supernova (SN) cooling~\cite{Fayet:2006sa,Bertoni:2014mva}, or during their transit to Earth across typically $\mathcal{O}$(10) kpc distances, they can scatter with the halo DM or relic background and deviate from the supernova (SN) line of sight, so that their flux is effectively attenuated; however, from scatters on DM away from the SN line of sight, there may be time-delayed ``echoes''; further, historic SNe and DM--$\nu$ scatters could form a diffuse Galactic background of neutrinos~\cite{Kolb1987,Scholberg2012,Murase2019,Carpio2023,Dev2025,Chauhan_2025}.
The scattering of supernova neutrinos (SN$\nu$) on DM may also alter the density profile of dwarf spheroidals~\cite{Heston:2024ljf}.
Moreover, event spectra of the integrated neutrino emission from all past CCSNe across cosmic history giving rise to the diffuse supernova neutrino background (DSNB), recently seen at the 2.6~$\sigma$ level at Super-Kamiokande loaded with gadolinium (SK--Gd)~\cite{SKDSNBhint} and expected to be observed at Hyper-Kamiokande, JUNO, and DUNE~\cite{Koshio:2025fjs,JUNOdsnb:2022lpc,Moller:2018kpn}, may carry information about interactions of the DSNB $\nu$s with DM~\cite{FarzanPalomares2014,Balantekin:2023jlg,Tseng:2024akh}. 
 
Even prior to the onset of core collapse, massive stars in their final evolutionary stages -- particularly when burning silicon -- emit a steady flux of MeV-energy neutrinos created by $e^+e^-$ pair annihilation, photo-neutrino production, nuclear $e^\pm$ capture, and $\beta^\pm$ and plasmon decays. 
These pre-supernova neutrinos (preSN$\nu$), though less luminous than supernova neutrinos (SN$\nu$) by a few orders of magnitude, carry direct information about the internal state of the dying star~\cite{kato2020theoretical}, and could provide an early warning of an impending Galactic supernova by hours to days~\cite{SNEWS:2020tbu}; some 30 massive stars in the solar neighborhood have been identified as progenitor candidates~\cite{Mukhopadhyay:2020ubs}, such as $\alpha$~Orionis, a.k.a. Betelgeuse. 
Due to their low fluxes and energies, their observation requires detectors with large target masses and low energy thresholds~\cite{asakura2016kamland,kato2017neutrino,Raj_2020,simpson2019sensitivity,li2020prospects,Super-Kamiokande:2022bwp,machado2022pre,abusleme2024real}.
In this work we estimate the sensitivities of such detectors to DM--$\nu$ interactions from the attenuation of the preSN$\nu$ fluxes from a future nearby SN event, typically $\mathcal{O}(100)$ pc away.  
In particular, we derive upper limits on the reduced cross section of elastic DM--$\nu$ scatters from estimated event counts of electron anti-neutrinos in the inverse beta decay (IBD) detection channel at Super-Kamiokande, KamLAND, and JUNO.
In forthcoming work, we derive similar sensitivities from attenuation and time-delay signals from (the usually studied post-bounce) neutrinos from a future Galactic supernova event~\cite{TallurRajSNnu}.
To our knowledge, ours is the first work to use preSN$\nu$ to study physics beyond the SM. 
[Ref.~\cite{mori2022presupernova} studied ultra-light axions produced in the final evolutionary stages of nearby SN progenitors from the conversion of MeV {\em photons}.]
Though we will have to be uncommonly fortunate for the occurrence of the next Galactic supernova -- in itself a rare event -- to be within the 100~pc neighborhood, the distinct characteristics of preSN$\nu$ make studying their potential in cornering new fundamental physics worthwhile.

Due to typically smaller DM column densities and lower statistics, our preSN$\nu$ sensitivities would appear weaker than those obtained from post-bounce SN$\nu$.
However, they are still of fundamental value:
(i) as these are $\mathcal{O}$(MeV) neutrino energies as opposed to $\mathcal{O}(10~$MeV) SN$\nu$ energies, models with resonant scattering peaked near an MeV may be uniquely probed by preSN$\nu$. 
Further, scattering via light mediators would be infrared-enhanced and give more value to preSN$\nu$-based sensitivities;
(ii) they would provide an important cross-check to the sensitivities derivable from the SN$\nu$ that would emerge from the same site, and moreover the ratio of the cross section limits would give us a clue as to the energy dependence of the cross section;
the slower, hours--days evolution of the preSN$\nu$ flux compared to the 10 second evolution of the SN$\nu$ flux could in principle also serve to probe the energy dependence of the DM scattering: a cross section that rises or falls with neutrino energy would distort/tilt the observed spectrum in a specific way;
(iii) due to sub-kpc baselines, the preSN$\nu$-based limits depend only on the local DM density, which is relatively well-measured with a fractional uncertainty of about 100\%. 
In contrast, for the $\mathcal{O}(10~$kpc) baselines of SN$\nu$, the fractional uncertainty on the DM column density can be much higher as inherited from fits of the Galactic DM density profile to stellar rotation curves; 
(iv) on a related note: in models of dissipative dark matter, a dark disk closely aligned with the baryon disk of the Milky Way may be formed~\cite{Fan:2013yva,darkdisk:WinchSetfordCurtin:2020cju}; 
it is plausible that the dissipative component of DM, with its rich interaction structure, is just the one that connects to the neutrino sector.
In such a scenario, the DM column density of SN$\nu$ is provided by the dark disk alone, weakening the SN$\nu$-based limits and hence lifting the significance of preSN$\nu$;
(v) the modeling of preSN$\nu$ and SN$\nu$ fluxes are of a different nature: 
the preSN $\bar\nu_e$ spectrum is set by $e^+e^-$ annihilation emissivity, in turn functions of the progenitor core's temperature, density and free electron fraction profiles, as tracked by stellar evolution codes, giving a quasi-thermal spectral shape.
For SN$\nu$, there are much larger theoretical systematics: explosion dynamics, standing accretion shock instability effects, equation of state-dependent average neutrino energy and pinching parameter, collective neutrino oscillations, phase transitions at accretion/explosion, and other actively debated effects.
These again make preSN$\nu$ sensitivities an important independent probe of new physics;
(vi) in case the progenitor undergoes a ``failed'' supernova event that collapses to a black hole soon after bounce, SN$\nu$ may be in shortage, and the preSN$\nu$ flux may carry much constraining power;

(vii) Further, as we will discuss, there may be an interesting interplay between (pre)SN$\nu$-based limits and those from cosmology. 
Specifically, there is a 3$\sigma$ level hint seen in Lyman-alpha data of DM-$\nu$ scattering suppressing the small-scale matter power spectrum~\cite{Hooper:2021rjc}.
We show that preSN$\nu$ may test this finding for DM-$\nu$ scattering cross sections that are a steeply increasing function of neutrino energy.

\begin{figure*}[ht]
    \centering
    \begin{minipage}{0.5\textwidth}
        \centering
        \includegraphics[width=\linewidth]{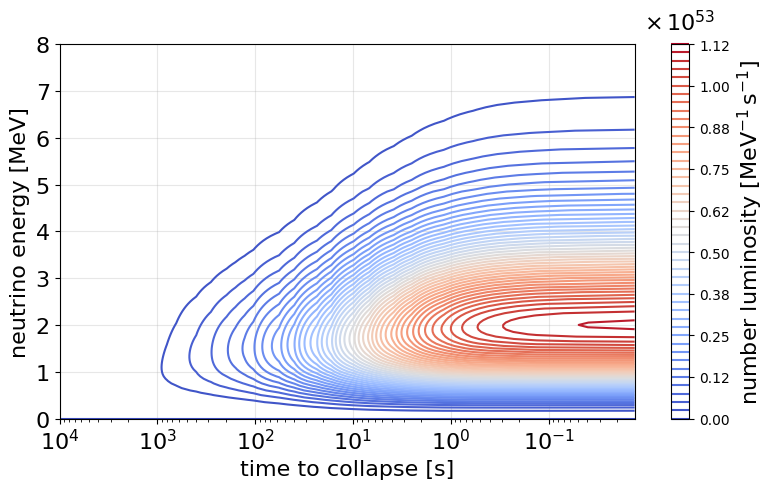}
    \end{minipage}
    \hfill
    \begin{minipage}{0.45\textwidth}
        \centering
        \includegraphics[width=\linewidth]{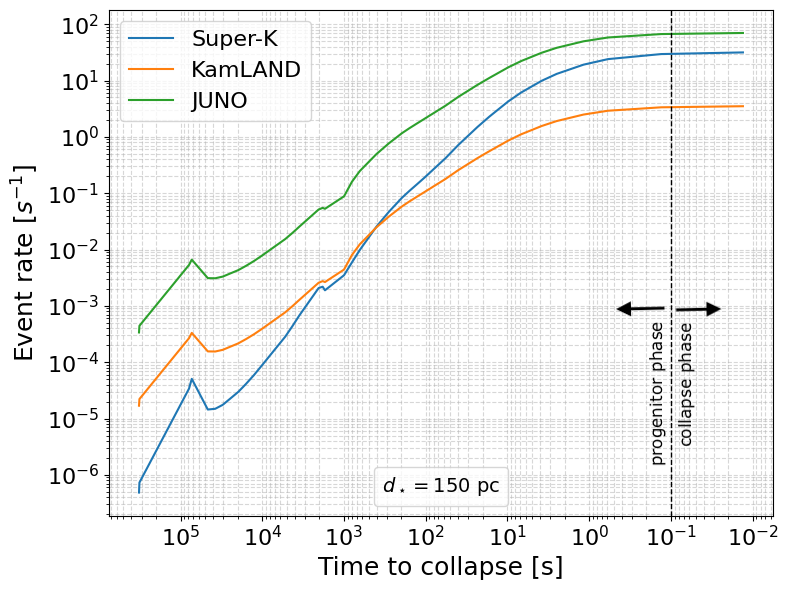}
    \end{minipage}
    \caption{{\bf \em Left.} Contours of pre-supernova electron anti-neutrino flux across neutrino energies and time before core collapse. 
    {\bf \em Right.} The corresponding event rates as obtained from Eq.~\eqref{eq:eventrate}. 
    The dashed vertical line marks the onset of the collapse phase, during which the flux is modeled as a Fermi-Dirac thermal spectrum.
     See Sec.~\ref{sec:setup} for further details.
    }
    \label{fig:fluxeseventrate}
\end{figure*}

\begin{table*}[ht]
    \centering
    \begin{tabular}{lccccccc}
        \hline
        detector &  $N_{\rm target}$ & $E_{\rm th}$ (MeV) & $N_{\rm tot}$ &  $\epsilon_{\rm stat}(\%)$ & $\epsilon_{\rm syst}(\%)$ & $N^{\rm CCSN}_{\rm tot}$ & limit on $\sigma_{\chi\nu}/m_\chi$ (10$^{-21}$ cm$^2$/GeV)  \\
        & & & & & & & \{$\tau_\chi^{2\sigma}$ ($<$~0.1~s), $\tau_\chi^{2\sigma}$ ($>$~0.1~s), $\tau_\chi$=0.1, $\tau_\chi$=1\} \\
        \hline
        Super-K &  $2.14 \times 10^{33}$ & 5.3 & 189 &  7.3 & 2.2 & $6.3 \times 10^7$ & \{$ 1.0, 8.8, 0.54, 5.4$\} \\
        KamLAND  &  $8.47 \times 10^{31}$ & 1.8 & 27  & 19.4 & 1.8 & $1.3\times 10^6$ & \{$2.8, 26, 0.54, 5.4$\} \\
        JUNO &  $1.69 \times 10^{33}$ & 1.8 & 534 & 4.3 & 1.0 & $2.5 \times 10^8$ & \{$0.6, 5.9, 0.54, 5.4$\} \\
        \hline
    \end{tabular}
     \caption{Detectors considered in this work with parameters taken from Ref.~\cite{kato2017neutrino}.
     The number of preSN$\nu$ events $N_{\rm tot}$ from a progenitor at $d_\star$ = 150~pc is computed using Eq.~\eqref{eq:eventrate}, and the CCSN events from the same progenitor $N_{\rm tot}^{\rm CCSN}$ using SNOwGLoBES.
     The last column provides our sensitivities on the reduced cross section by the criteria described in Sec.~\ref{sec:results}.
}
    \label{tab:detectors}
\end{table*}

This paper is organized as follows. 
In Section~\ref{sec:setup} we present the preSN$\nu$ flux and detector event rates, and discuss the attenuation of the flux by scattering on dark matter to derive a criterion for setting limits.
In Section~\ref{sec:results} we show and discuss our results.
In Section~\ref{sec:discs} we provide further discussion and conclude.

\section{Setup} 
\label{sec:setup}

In this section we sketch the signals of DM--preSN$\nu$ interactions.
From preSN$\nu$ spectra available online we first compute the expected event counts at various detectors in relevant detection channels. 
We then describe our procedure for setting limits on DM--$\nu$ scattering via attenuation of the preSN$\nu$ and SN$\nu$ fluxes.

\subsection{Pre-supernova neutrino flux, event rate at detectors}

Using stellar evolution codes preSN$\nu$ emission rates have been estimated with increasing sophistication~\cite{Odrzywolek:2004em,Patton:2015sqt,Kato:2015faa,Yoshida:2016imf,kato2017neutrino}, including a recent claim that accounting for thermally excited nuclei could enhance $\nu$ emission by an order of magnitude~\cite{Dzhioev:2025dhs}.
In this study we will use the results of Ref.~\cite{kato2017neutrino}, tabulated in Ref.~\cite{kato_neutrino_2020}, while noting that the other flux estimates (that do not account for hot nuclei) result in sensitivities of the same order of magnitude.

We take preSN$\nu$ spectra from the $15M_{\odot}$ progenitor model in Ref.~\cite{kato_neutrino_2020}, from which we extract the differential number luminosity $d^2N_\nu/dE_\nu\,dt$ of $\bar\nu_e$s as a function of their energy $E_\nu$ and time before core collapse $t$. 
 To this end, we merged \texttt{spe\_all*} -- which provides the binned $\bar\nu_e$ spectrum at each simulation snapshot -- with the corresponding light-curve information from \texttt{lightcurve\_nueb\_all.dat}, which provides the physical time associated with each snapshot together with the integrated luminosity.
 By matching the timestep labels in \texttt{step.dat} with the entries in the light-curve file, we assigned a co-ordinate $t$ to each spectrum.

The left panel of Fig.~\ref{fig:fluxeseventrate} shows contours of constant $d^2N_\nu/dE_\nu\,dt$ of pre-supernova $\bar\nu_e$s in $E_\nu$-$t$ space. 
As the system evolves toward core collapse (going from left to right), the flux increases significantly due to enhanced weak interaction rates, with the increasing density of contours near collapse marking rapid evolution of the $\bar\nu_e$ emission. 
The emission is mainly at around 2~MeV, although the spectrum broadens and extends to $\sim5-7$~MeV closer to collapse. 
The higher-energy tail is particularly relevant for detection at Super-K with its steep energy threshold.

The preSN$\nu$ event rate at a detector, and the number of events detected within a time window [$t_{\rm i}, t_{\rm f}$], are given by
\begin{eqnarray}
\nonumber     \frac{dN_{\rm event}}{dt} &=& \frac{N_{\rm target}}{4\pi d_\star^2} 
    \int_{E_{\rm th}}^{\infty} \sigma_{\nu T}(E_\nu)\, \frac{d^2N_\nu}{dE_\nu\,dt}\, dE_\nu~, \\
     N_{\rm tot} &=& \int_{t_{\rm i}}^{t_{\rm f}} \frac{dN_{\rm event}}{dt}\, dt~,
    \label{eq:eventrate}
\end{eqnarray}
where $d_\star$ is the progenitor distance from Earth,
$N_{\rm target}$ is the number of target nuclei, 
$\sigma_{\nu T}(E_\nu)$ is the interaction cross section of the neutrino and target nucleus, and 
$E_{\rm th}$ is the detector energy threshold.
While Super-K, KamLAND and JUNO are suitable for different sets of detection channels~\cite{kato2020theoretical}, the primary channel in all three is IBD ($\bar{\nu}_e + p \rightarrow n + e^+$), for which the cross section is~\cite{beacomvogel1999}
\begin{equation}
    \sigma_{\nu T}(E_\nu) = 9.52 \times 10^{-44}\,\text{cm}^2 \bigg( \frac{E_e}{\rm MeV}\bigg)\, \bigg(\frac{p_e}{\rm MeV}\bigg)~,
\end{equation}
where $E_e = E_\nu - 1.293$ MeV and 
$p_e = \sqrt{E_e^2 - m_e^2}$ are respectively the positron energy and momentum. 

In this work we compute $N_{\rm tot}$, which determines our limits, for a 15$M_\odot$ progenitor, with normal neutrino mass ordering as treated in Refs.~\cite{kato2017neutrino,kato_neutrino_2020}.
While this is a typical progenitor mass, e.g., consistent with that of Betelgeuse's~\cite{Neuhauser2022}, these works also provide fluxes for 12$M_\odot$ and 9$M_\odot$ progenitors.
 The former emits $\mathcal{O}(1)$ fewer neutrinos of all flavors than the 15$M_\odot$ progenitor, and the latter, being an (uncommon) electron-capture SN, emits much fewer $\bar{\nu}_e$s and much more $\nu_e$s.
They also provide fluxes for the inverted mass ordering, which are also smaller.
On the whole, these variations may weaken our limits by up to a factor of 3, which (as we mention in Sec.~\ref{sec:results}) may be partially compensated by the uncertainty in the local DM density.
In Table~\ref{tab:detectors} we summarize the event rates at Super-K, KamLAND and JUNO -- detectors highlighted in the review Ref.~\cite{kato2020theoretical} -- with relevant parameters taken from Ref.~\cite{kato2017neutrino}, with whose event counts our computed values are consistent. 
In principle future experiments such as Hyper-K and DUNE (in the $\nu_e$ channel) may also detect preSN$\nu$, however the Hyper-K event count is expected to fall within that of the range of the detectors we consider, and DUNE with high thresholds is expected to see fewer events~\cite{kato2017neutrino}, and we do not consider them further.
The backgrounds at the detectors we consider are negligible, particularly when reactor neutrino fluxes are small and when Gd is loaded into Super-K~\cite{kato2017neutrino,li2020prospects,Super-Kamiokande:2022bwp}.

The right panel of Fig.~\ref{fig:fluxeseventrate} shows the time evolution of the predicted $\bar\nu_e$ event rate in different detectors as a progenitor at $d_\star = 150~$pc approaches core collapse. 
As expected from the evolution of the flux in the left panel, the rates are suppressed at early times, then rise steadily and accelerate sharply during the final hours and minutes before collapse. 
The separation between the curves becomes most pronounced close to collapse, where the neutrino luminosity is highest and the accumulated signal grows rapidly. 
In the progenitor phase ($t > 0.1$ s), the event rate increases steadily over several orders of magnitude as the core contracts and the temperature rises, enhancing both the neutrino luminosity and average energy. The dips visible around $t \sim 10^4$--$10^5$ s correspond to temporary suppression in the neutrino emission caused by the ignition of O and Si shell burnings, which expand the outer layers and reduce the central pressure and hence temperature momentarily~\cite{kato2017neutrino}. 
In the collapse phase ($t < 0.1$ s), the event rate plateaus since electrons become degenerate and block the phase space of $\beta^-$ decay, which is the dominant $\bar\nu_e$ production mode. 
In this phase the neutrino fluxes are modeled as Fermi-Dirac thermal spectra, as done for post-bounce SN$\nu$.
Despite being the largest detector, Super-K obtains the fewest events due to its Cerenkov-based threshold being higher than the typical preSN$\nu$ energies, while JUNO obtains the most events thanks to large mass and low threshold.


\subsection{Pre-supernova neutrino scattering on dark matter}
\label{subsec:preSNDM}

Our signal is a deficit of events due to preSN$\nu$ scattering on the intervening DM and deviating from the line of sight.
For the sub-kpc baselines for which at least one event is expected, the DM density $\rho_\chi$ is unvaried, which we take to be 0.4~GeV/cm$^3$.
Hence the optical depth for preSN$\nu$ scattering on DM of mass $m_\chi$ with cross section $\sigma_{\chi\nu}$ is given by
\begin{equation}
    \tau_\chi = \rho_\chi \left(\frac{\sigma_{\chi\nu}}{m_\chi}\right) d_\star.
    \label{eq:opticaldepth}
\end{equation}
We then estimate the 2$\sigma$ signal sensitivity by requiring a deficit that is twice the uncertainty in the number of events.
In practice, we set $\tau_\chi = 2~\times$~total fractional uncertainty. 
Thus, adding the statistical and systematic uncertainties in quadrature, the $\tau_\chi$ corresponding to 2$\sigma$ sensitivity is given by
\begin{equation}
    \tau_\chi^{2\sigma}
    =
    \frac{2\sqrt{N_{\rm tot} + (\epsilon_{\rm sys} N_{\rm tot})^2}}
    {N_{\rm tot}}\simeq\frac{2}{\sqrt{N_{\rm tot}}}~.
    \label{eq:opticaldepthuncert}
\end{equation}
The second (near-)equality comes from the systematic uncertainties being negligible compared to the statistical uncertainties, as seen in Table~\ref{tab:detectors}. 
The corresponding limit on the reduced dark matter--neutrino cross section is subsequently obtained as
\begin{equation}
    \left(\frac{\sigma_{\chi\nu}}{m_\chi}\right)\leq
    \frac{2}
    {\rho_\chi d_\star \sqrt{N_{\rm tot}}}.
    \label{eq:limitcriterion}
\end{equation}

We see from Eq.~\eqref{eq:eventrate} that, since $N_{\rm tot} \propto 1/d_\star^2$, the limit above is independent of the progenitor distance $d_\star$.

\subsection{Post-bounce neutrino scattering on dark matter}
\label{subsec:CCSNDM}

The progenitor treated in the previous subsections would explode and give CCSN $\nu$s, which also would scatter on DM on the way to Earth.
To show limits from the attenuation of this post-bounce flux, assuming the SN is successful (i.e., it doesn't collapse into a black hole), we obtain the number of $\bar{\nu}_e$ events (which would give direct comparisons to our preSN$\nu$-based limits) $N_{\rm tot}^{\rm CCSN}$ using the event generator SNOwGLoBES~\cite{snowglobes2021} with the ``Livermore'' flux parameterization~\cite{Totani:1997vj}.
These are listed in Table~\ref{tab:detectors}, from where we also take detector thresholds used in this computation.
To estimate our sensitivities we once again use the criterion that the optical depth is twice the total uncertainty (Eq.~\eqref{eq:limitcriterion}), except now we find that it is the systematic uncertainty that drives the limit as seen in Table~\ref{tab:detectors}.


\section{Results}
\label{sec:results}

\begin{figure}[t]
    \centering
    \includegraphics[width=0.45\textwidth]{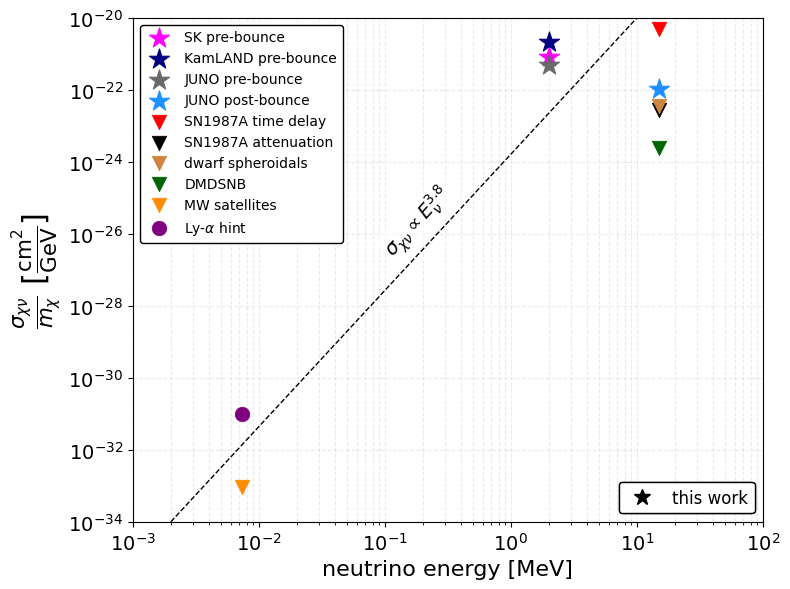}
    \caption{
    Sensitivities to the reduced cross section for DM-$\nu$ scattering from measurements of the pre-supernova neutrino flux centered around 2 MeV from a future nearby core-collapse supernova. 
    Also shown in different energy ranges are the sensitivity from post-bounce supernova neutrinos from the progenitor site,
    existing bounds from SN1987A $\nu$ flux measurements via the attenuation of the flux by DM and time-delay signatures by scattering with the DM halo~\cite{Chauhan_2025},
    the non-detection of a DM-diffused SN$\nu$ background (DMDSNB) flux from historic DM-SN$\nu$ scatters in the Galaxy~\cite{Chauhan_2025}, 
    the alteration of matter distribution in dwarf spheroidals~\cite{Heston:2024ljf},
    the suppression of small-scale structure constrained with the Milky Way satellite population~\cite{Crumrine:2024sdn},
    and a hint of DM-$\nu$ interactions in Lyman-$\alpha$ data on the matter power spectrum~\cite{Hooper:2021rjc}.
    See Sec.~\ref{sec:results} for further details.
    }
    \label{fig:limits}
\end{figure}

In the last column of Table~\ref{tab:detectors} we specify four sets of limits for the 15$M_\odot$ progenitor model: 
those obtained from $\tau_\chi^{2\sigma}$ accounting for events in the collapse phase ($t < 0.1$~s), 
the same in the progenitor phase ($t > 0.1$~s),
those obtained by setting $\tau_\chi = 0.1$ in Eq.~\eqref{eq:opticaldepth}, corresponding to a 10\% DM--$\bar\nu_e$ scattering probability, and
those from $\tau_\chi = 1$, corresponding to a conservative criterion by which all $\bar{\nu}_e$ are scattered by DM.
The $\tau_\chi^{2\sigma}$-based limits in the two phases of neutrino emission apply in two different regimes: in the progenitor (collapse) phase, the flux is obtained numerically (modeled as a Fermi-Dirac thermal spectrum).
Clearly, the overall limit is driven by the collapse phase, when the flux is highest.
Among the three detectors, KamLAND yields a slightly weaker limit from the $\tau_\chi^{2\sigma}$ criterion, primarily due to  
its smaller target mass and hence fewer detected events. 

Figure~\ref{fig:limits} shows our sensitivities from the criterion in Eq.~\eqref{eq:limitcriterion}.
As seen from Fig.~\ref{fig:fluxeseventrate} and Table~\ref{tab:detectors}, these are driven by events from the collapse phase.
As explained in Sec.~\ref{subsec:preSNDM}, these limits are insensitive to the progenitor distance.
While the limits we show are for the 15$M_\odot$ progenitor with normal hierarchy of neutrino masses as modeled in Ref.~\cite{kato2017neutrino}, they may weaken by up to a factor of 3 from variation in progenitor mass and mass hierarchy; however, the factor-of-2 uncertainty in determinations of the local DM density~\cite{deSalas:2020hbh} may partially make up for this.

Also shown are limits in 10$-$20~MeV energies based on CCSNe.
This includes the sensitivity from the explosion of the very progenitor from which we obtain the preSN$\nu$ limits, using the criterion discussed in Sec.~\ref{subsec:CCSNDM}; we only show the strongest sensitivity, which comes from JUNO.
While these are at generally smaller $\sigma_{\chi\nu}/m_\chi$ than our sensitivities, they operate in an energy range different from ours, implying that in scenarios with resonant scattering peaked near an MeV, or with light mediators enhancing soft scatters, ours may be a unique probe of present-day DM$-\nu$ scattering.
As also mentioned in the Introduction, in the case of a positive detection, the ratio of inferred cross sections would suggest its $E_\nu$-dependence.
For instance, if $\sigma_{\chi\nu}$ is energy-independent, the ratio is 1; 
for $m_\chi \ll E_\nu$,  if $\sigma_{\chi\nu} \propto s = m_\chi^2 + 2E_\nu m_\chi \simeq 2E_\nu m_\chi$, it would be $\sim$1/5, and if $\sigma_{\chi\nu} \propto 1/s$ it would be $\sim$5.
Also as mentioned, our limits could prove handy as a cross-check to SN$\nu$-based limits, due to qualitative differences in the modeling of preSN$\nu$ vs SN$\nu$ fluxes and fits to local vs Galactic DM densities, or if the CCSN fails too soon or in dark disk scenarios.

Figure~\ref{fig:limits} also shows at around $5-10$~keV a limit from structure formation affecting Milky Way satellite counts and a parametric point favored by Lyman-alpha observations of the matter power spectrum.
The energy range in question is set by the structure-forming perturbation modes that re-enter the horizon, $k \sim 1-10~h$/Mpc.
Interestingly, this means the Lyman-$\alpha$ hint may be tested by preSN$\nu$ measurements for certain DM-$\nu$ interaction models.
We show this in Fig.~\ref{fig:limits} for a toy scenario with $\sigma_{\chi\nu} \propto E_\nu^{3.8}$, which is allowed by the hint and evades all other limits except those from future preSN$\nu$.
While this is not a realistic model, it again illustrates the importance of looking for DM-$\nu$ interactions in all energy ranges possible.

\section{Discussion}
\label{sec:discs}

In this work we used the distinct virtues of pre-supernova neutrinos to probe new physics in the form of dark matter-neutrino scattering interactions, with results summarized in Fig.~\ref{fig:limits}.
Due to the unique combination of their detectably large fluxes, astronomical baselines and  $\mathcal{O}$(MeV) energies, preSN$\nu$ can probe certain energy dependences in DM-$\nu$ scattering and may help test hints of this process in Ly-$\alpha$ spectral data.
Other scenarios of new physics too may be probed by preSN$\nu$, e.g., their scattering on cosmic relics such as the cosmic neutrino background or a background of vectors or majorons that couple to $\nu$s~\cite{Kolb1987}; and the invisible decay of neutrinos, with measurements of the preSN$\nu$ flux providing an independent check of the future limits from a Galactic CCSN or the DSNB~\cite{Martinez-Mirave:2024hfd}.
Beyond the currently operational detectors we had considered, one could also estimate the sensitivities of DM-preSN$\nu$ scattering in future detectors such as DUNE and Hyper-K~\cite{kato2017neutrino}, and those employing coherent elastic nuclear scattering channels~\cite{Raj_2020,Lang:2016zhv,Raj:2019sci,Pattavina:2020cqc}, as well as at proposals for kilotonne-scale low-threshold detectors such as 
a dual-phase 10~kt ``module of opportunity'' at DUNE~\cite{DUNEModuleDM:PNL2020,DUNEModuleDM:snowmass:Avasthi2022,DUNEModuleDM:Bezerra2023}, or liquid xenon-based kilotonne detectors designed for $0\nu\beta\beta$ searches repurposed for DM searches~\cite{Xekton:Avasthi:2021lgy,Xekton:Anker:2024xfz}.
We leave these interesting investigations to future studies.

\section*{Acknowledgments}

We appreciate thoughtful suggestions by
Ranjan Laha,
Aadarsh Singh, and 
Anand Sharma. 
N. R. thanks IISER Berhampur for their hospitality during {\em First Few Seconds 2026}, where this work was initiated.
N. R. acknowledges support from the grant ANRF/ECRG/2024/000387/PMS and the Infosys Foundation, Bangalore.

\newpage
\bibliography{refs}
\end{document}